%% file: main.tex
\documentclass[sigplan,screen]{acmart}

\usepackage[utf8]{inputenc} 
\usepackage[T1]{fontenc}    
\usepackage{hyperref}       
\usepackage{url}            
\usepackage{booktabs}       
\usepackage{amsfonts}       
\usepackage{nicefrac}       
\usepackage{microtype}      
\usepackage{cleveref}       
\usepackage{lipsum}         
\usepackage{graphicx}
\usepackage{natbib}
\usepackage{doi}
\usepackage{xcolor}
\usepackage{amsmath}
\usepackage[ruled,vlined]{algorithm2e} 
\usepackage{subcaption}
\usepackage{tablefootnote}
\usepackage{balance}

\title{Linker Code Size Optimization for Native Mobile Applications}

\author{Gai Liu}
\orcid{https://orcid.org/0000-0001-8538-686X}
\affiliation{%
  \institution{ByteDance}
  \city{Mountain View, CA}
  \country{USA}
}
\email{gai.liu@bytedance.com}

\author{Umar Farooq}
\orcid{https://orcid.org/0000-0001-7229-9847}
\affiliation{%
  \institution{ByteDance}
  \city{Mountain View, CA}
  \country{USA}
}
\email{umarfarooq@bytedance.com}

\author{Chengyan Zhao}
\orcid{https://orcid.org/0000-0002-2054-8677}
\affiliation{%
  \institution{ByteDance}
  \city{Mountain View, CA}
  \country{USA}
}
\email{chengyan.zhao@bytedance.com}

\author{Xia Liu}
\orcid{https://orcid.org/0000-0002-0546-9217}
\affiliation{%
  \institution{ByteDance}
  \city{Shenzhen}
  \country{China}
}
\email{liuxia.nathan@bytedance.com}

\author{Nian Sun}
\orcid{https://orcid.org/0000-0003-2762-4131}
\affiliation{%
  \institution{ByteDance}
  \city{Shanghai}
  \country{China}
}
\email{sunnian@bytedance.com}

\renewcommand{\shorttitle}{Linker Code Size Optimization for Native Mobile Applications}

\newcommand{\stitle}[1]{\vspace{1ex}\noindent\textup{\textbf{#1}}}

\newcommand{\TouTiao}{NewsFeedApp}
\newcommand{\TikTok}{ShortVideoApp}
\newcommand{\Lark}{CollaborationSuiteApp}

\hypersetup{
pdftitle={},
pdfsubject={},
pdfauthor={},
pdfkeywords={},
}

\setcopyright{acmlicensed}
\acmPrice{15.00}
\acmDOI{10.1145/3578360.3580256}
\acmYear{2023}
\copyrightyear{2023}
\acmSubmissionID{cc23main-p22-p}
\acmISBN{979-8-4007-0088-0/23/02}
\acmConference[CC '23]{Proceedings of the 32nd ACM SIGPLAN International Conference on Compiler Construction}{February 25--26, 2023}{Montréal, QC, Canada}
\acmBooktitle{Proceedings of the 32nd ACM SIGPLAN International Conference on Compiler Construction (CC '23), February 25--26, 2023, Montréal, QC, Canada}
\received{2022-11-10}
\received[accepted]{2022-12-19}

\begin{document}

\ccsdesc[500]{Software and its engineering~Compilers}
\ccsdesc[500]{Human-centered computing~Ubiquitous and mobile computing}

\keywords{Code Size Optimization, Static Analysis, iOS}

\begin{abstract}
Modern mobile applications have grown rapidly in binary size, which restricts user growth 
and hinders updates for existing users. 
Thus, reducing the binary size is important for application developers. 
Recent studies have shown the possibility of using link-time code size optimizations
by re-invoking certain compiler optimizations on the linked intermediate representation
of the program.
However, such methods often incur significant build time overhead and require
intrusive changes to the existing build pipeline.

In this paper, we propose several novel optimization techniques that do not require 
significant customization to the build pipeline and reduce binary size with 
low build time overhead. 
As opposed to re-invoking the compiler during link time, we perform true
linker optimization directly as optimization passes within the linker.
This enables more optimization opportunities such as pre-compiled libraries 
that prior work often could not optimize. 
We evaluate our techniques on several commercial iOS applications including 
\TouTiao{}, \TikTok{}, and \Lark{}%
, each with hundreds of millions of daily active users. 
Our techniques on average achieve 18.4\% binary size reduction across the three commercial
applications without any user-perceivable performance degradations. 


%

\end{abstract}

\maketitle


\input{intro}

\input{technique}

\input{experiment}

\input{related}

\input{conclusion}

\balance

\bibliographystyle{unsrtnat}
\bibliography{references}  

%
%
%

\end{document}

%% file: intro.tex
\section{Introduction}
\label{sec:intro}
Mobile applications have seen tremendous adoption over the last decade. 
Today, billions of users depend on them for a variety of reasons, 
including access to news, social media, ride sharing, work productivity, and much more. 
In this competitive and vastly growing environment, constantly delivering new 
features is of prime importance to application developers. 
However, the proliferation of new features results in a huge increase in binary size~\cite{app-size-blog, superpack-fb}. 
At the same time, mobile devices provide limited storage, and distribution 
channels (i.e., app stores) enforce download-size restrictions. For example, 
the Apple App Store~\cite{app-store} requires a Wi-Fi connection to download applications 
larger than 200 MB as of 2020. 
This download-size restriction by the app store limits application growth as new 
installations and updates, including security improvement, cannot be performed without a Wi-Fi connection.

\stitle{Code size optimization.} 
Compiler optimizations are effective in minimizing the size of the compiled binary~\cite{survey-code-size-reduction}.
In addition to performance benefits, many compiler optimizations may also reduce code size, such as dead and unreachable code elimination~\cite{torczon2007engineering}, common sub-expression elimination~\cite{gcse1970}, partial redundancy elimination~\cite{kennedy1999}, 
constant and copy propagation~\cite{wegman1991,aho2007compilers}, constant folding~\cite{Rodriguez2016}, value numbering~\cite{rosen1988global}, register allocation and instruction scheduling~\cite{lau2003}, 
code compression~\cite{ernst1997,edler2014,chen2003code}, and peephole optimizations~\cite{massalin1987superoptimizer}. 
Other than compiler optimizations, link-time optimizations (LTO) and 
post-link-time optimizations~\cite{lto1,caldwell2017reducing,debray2000compiler,schwarz2001plto,de2007link,chanet2005system,he2007code,pitre-lwn,panchenko2019bolt,llvm-propeller} 
have also shown success in reducing code size.

\stitle{State of the art.} 
iOS applications are commonly compiled using LLVM~\cite{lattner2004llvm},
and several of the above-mentioned optimizations are available in LLVM by default.
One of the key size optimization passes is the machine outlining pass that extracts
frequent sequences of instructions into separate functions at the machine intermediate representation (IR) level
to reduce the code size~\cite{llvm-machine-outline}.
The machine outlining pass is scheduled as part of the compilation pipeline, 
which operates on a single compilation unit during compilation.
Optimizing only within a single compilation unit leaves much room for size reduction 
on large applications.
This is because a significant portion of the repetitive patterns is common across
multiple files/compilation units.

\begin{table}[t!]
\centering
\small
\caption{Summary of representative size optimization approaches for native mobile applications. Build time overhead is the increase in build time when enabling the corresponding techniques.}
\begin{tabular}{lccc}
\toprule

Approach      					        & Build time & Require custom & Binary  \\
              					        & overhead   & build pipeline & size saving    \\ 
\midrule
Chabbi et al.~\cite{uber-lto} 	& 200\%\tablefootnote{This technique triples the build time according to~\cite{uber-lto}.}   & Yes             & 17.6\%           \\
Lee et al.~\cite{fb-pgo}    	  & 40\%    & Yes            & 12.6\%           \\
This work     					        & 17\%    & No             & 18.4\%      	    \\
\bottomrule
\end{tabular}
\label{tbl:comparsion}
\end{table}


To enable global size optimizations across compilation units, 
state-of-the-art approaches often utilize recent development in LTO~\cite{lto1, johnson-lto}.
Chabbi et al.~\cite{uber-lto} proposed a whole-program machine code outlining using full LTO (also known as monolithic LTO).
The full LTO process merges all the LLVM IR files generated from the input files into a single IR module, and applies transformation and lowering passes on the merged module.
While this approach provides a significant binary size reduction, it contributes up to 45 minutes overhead in the build pipeline since the merged module can only be processed by a single thread.
This overhead triples the overall build time~\cite{uber-lto}.
Additionally, this approach requires the source files or the IRs to be available for optimization\footnote{Embedding LLVM IR in the object files requires explicitly specifying \texttt{-fembed-bitcode} option, which is off by default in Clang.}, which means binary-only third-party libraries cannot be effectively optimized.
On top of this, a customized build pipeline for linking the IRs needs to be built,
which incurs non-trivial changes to the existing flow~\cite{uber-lto}.

More recently, Lee et al.~\cite{fb-pgo} extended the machine outlining pass and 
modifies the LLVM compilation pipeline to enable global machine outlining. 
The authors proposed to run the code generation step twice.
In the first iteration, they traverse each file to collect optimization opportunities.
In the second round, actual optimizations are performed.
Their approach requires significant changes to the compilation pipeline and the
two-round code generation incurs around 40\% build time overhead.
Since their approach works at the IR level, source files would be needed 
to be optimized.
Collectively, these efforts introduce effective code size optimizations
for iOS applications. 
Nonetheless, eliminating the requirement of a custom build pipeline, 
reducing the build time overhead, and extending the optimizations to third-party
libraries remain the challenges to solve.
We compare our approach and other existing techniques in Table~\ref{tbl:comparsion}, where the binary size saving numbers are taken from the apps with the largest size reductions in \cite{uber-lto} and \cite{fb-pgo}, respectively.

\stitle{Overview of this work.} 
To address these challenges, we propose a novel framework to perform linker optimization%
\footnote{In this work, link time optimization denotes the existing approach of 
invoking the optimizer during linking through a shared object such as \texttt{libLTO}~\cite{llvm-lto}. 
Linker optimization refers to our approach of directly implementing optimization passes in the linker.}
to reduce the binary size, which does not require customizing the build pipeline 
and the time overhead remains within 17\% of the overall build time. 
We extend the open-source \texttt{ld64} linker~\cite{ld64} with
additional analyses and size-optimizing transformation passes.
This enables the build pipeline to leverage the optimizations by simply using our customized linker
without needing to change any existing compiler/linker flags.
Specifically, our instruction decoding utilities allow the linker to understand 
the semantics of the machine instructions. 
Our extension to function hashing enables efficient identity check.
The instruction visibility analysis ensures safe code transformations that maintain
the correct semantics of the program and its associated metadata.
These analyses enable us to perform code size optimizations, including general
sequence outlining, frame code outlining, and identical code folding.
We evaluate our linker on three widely used commercial mobile applications in terms of 
code size reduction and build time overhead. 
Our evaluations show that our technique reduces the binary size of \TouTiao{}, 
\TikTok{}, and \Lark{} by 17.8\%, 20.5\%, and 17.1\%, respectively.

%
%
%
%

In this work, we make the following contributions:
\begin{itemize}
  \item To the best of our knowledge, we are the first to propose conducting code size optimization within the linker, as opposed to the existing approach of piggybacking on the compiler's optimization passes.
  \item We describe a novel framework within the iOS linker for code size optimization including the necessary analyses and code transformations.
  \item We show that our techniques achieve best-in-class results in both code size and build time on several real-world iOS applications without user-noticeable performance degradations.
\end{itemize}

\input{figs/overview}

Next, we describe our techniques, including analyses and optimizations
in Section~\ref{sec:technique}. 
We discuss implementation details in Section~\ref{sec:implementation}, 
then we present evaluations and experiments in 
Section~\ref{sec:experiment}, followed by the related work in Section~\ref{sec:related}. 
Finally, we conclude this work in Section~\ref{sec:conclusion}.

%% file: figs/overview.tex
\begin{figure*}[h]
       \centering
       \includegraphics[width=0.95\textwidth]{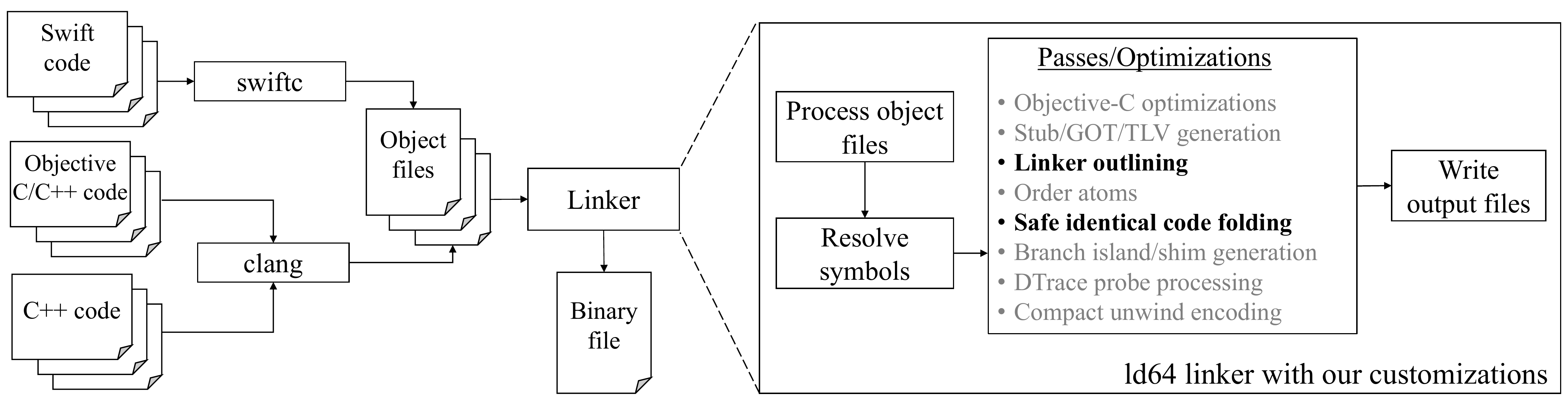}
       \caption{Overview of the build pipeline and detailed view of the \texttt{ld64} linker with our new passes highlighted.}
       \label{fig:overview}
\end{figure*}

%% file: technique.tex
\section{Techniques}
\label{sec:technique}

In this section, we describe the main techniques to reduce the code size of native iOS applications.
We first discuss the common analyses that we use across the optimizations, then
present the specific optimizations that effectively reduce the code size.
We propose two main types of code size reduction techniques,
(i) linker outlining aims at finding repetitive sequences of instructions within
functions and then outlining them into a shared function, and 
(ii) safe identical code folding explores repetitive functions and merges them into 
a single function while ensuring the correctness under function pointer comparisons.
Figure~\ref{fig:overview} shows the general build pipeline, and we highlight our contributions to the
linker in the flow.

\subsection{Analyses}
\label{sec:analyses}

To enable code transformations during linking, we develop a few useful
analyses on the machine instruction level. This enables us to analyze the function-level
and instruction-level properties and make optimization decisions.

\stitle{Instruction decoding utilities.}
Traditional linkers do not have the full instruction decoding capability
since they do not need to know the fine details of all instructions.
However, in our work, it is critical that we fully decode the instructions 
and unveil details including opcode, immediate values, register usage, and any
special mode flags.
We construct a comprehensive set of utility functions in the \texttt{ld64} linker for 
the AArch64 ISA~\cite{arm64-arch} and use them throughout the analyses and optimizations.
The utility functions resemble functionalities that are commonly found in a typical compiler
where information such as the type, opcode, and register indices of an instruction is obtained
from its binary encoding.

\stitle{Function hashing.}
Hashing a function is a common routine used in our optimizations. It enables
efficient identity checks across multiple functions, which is an important operation
in code folding.
Here, we extend the existing function hashing utility in \texttt{ld64}.
In our scheme, we hash each function into a 64-bit unsigned integer. 
Specifically, the hash of a function is determined by two factors. First, the 
machine instructions are iteratively hashed with a prime multiplier. 
Second, the metadata (fixups in the \texttt{ld64} context) is hashed from strings 
to integers and added to the hash value from the first step.
This two-step hashing scheme ensures control flow information such as branch
targets are encoded in the computed hash.

The necessary condition for two functions to be identical is that their hash
values are identical.
However, having identical hash values is not sufficient to prove that two functions
are identical due to potential hash collisions.
Bytewise comparison is needed to prove sufficiency.
To this end, we implemented an optional safety check in the pass to conduct byte-wise 
identity check across all functions with the same hash value.
Since the number of functions mapping to the same hash value is relatively small 
on average compared to the total number of functions in the program, the compile
time overhead of this safety check is small.

\input{figs/instruction_visibility}

\stitle{Instruction visibility.}
Since our optimizations are conducted during the late stage of the build pipeline,
we need to be aware not to optimize away certain instructions that other instructions or metadata may explicitly reference.
We define \textit{visible} instructions to be the branch targets of control flow instructions 
and the start/end/jump targets in an exception handling table.
Removing such visible instructions could lead to unexpected application behaviors
such as incorrect logic and inaccurate exception handling.
To understand their semantics and identify these special instructions, we linearly scan the instruction sequence
and parse the exception handling tables embedded in the object files.
Information related to the visible instructions is preserved upon discovery and reused by various optimization passes during linking. 
Figure~\ref{fig:instruction_visibility} shows an example of visible instructions. 
The two highlighted \texttt{ldr} instructions are the branch targets of their
corresponding \texttt{cbz}/\texttt{cbnz} instructions.
Removing or outlining these instructions may lead to incorrect behaviors, thus we consider
them visible to other instructions and skip them during outlining.

\subsection{General Sequence Outlining}
\label{sec-general-outline}

Outlining is a key technique in reducing the code size.
Outlining discovers common code sequences in a program and replaces them with
calls to the corresponding outlined sequences.
Traditionally, outlining is commonly done during compilation. For example,
LLVM employs a machine-outliner pass that operates on the machine IR.
When combined with full LTO, LLVM-based outlining can explore optimization 
opportunities at the whole program granularity~\cite{uber-lto}.
We instead propose and implement outlining inside a linker for the following
reasons.
First, mobile applications are usually written in a modularized fashion, and 
it is common to include third-party libraries which are pre-compiled into
object code. This development flow is not compatible with the LLVM-based outlining
solution since the source code of many modules is not available during build time.
Second, full LTO based whole program outlining significantly increases the build time 
by tens of minutes to a few hours~\cite{uber-lto}.
Such a long compile time poses a significant challenge for integration into rapid development pipelines.
Third, full LTO based build flow complicates incremental compilation and
makes incremental debugging significantly more difficult.

We develop a general code sequence outliner in the \texttt{ld64} linker that can optimize the whole program. 
For example, our linker outliner can optimize third-party libraries available only in binary
format, which is beyond the capability of the LLVM machine outliner.
Algorithm~\ref{alg-general-outline} details the steps in our general sequence outliner.
Our outliner first traverses the entire program and hashes every instruction sequence
whose length is within a predefined range (e.g., from length-2 to length-12). 
We hash the instruction sequences by extending the function hashing technique introduced in Section~\ref{sec:analyses}
to handle arbitrary code sequences.
The range of sequence lengths is a user-configurable parameter, and we empirically
observe that sequences longer than 12 instructions are rarely repeated in our applications.
During the traversal, we keep track of the length of each hashed sequence,
and their occurrences efficiently by extensive hashing and caching.
We then employ a cost function to evaluate the profitability of a given sequence,
where both the length and the occurrences matter.
We provide linker flags to control the cost function's aggressiveness.
Next, we create the outlined functions and modify the control flow of the 
original code to branch to these outlined functions.
An outlined function either inherits the control flow from its original sequence
or returns control back to the caller once the outlined function finishes its execution.
Finally, we update the relevant metadata to reflect the changes due to outlining.
This includes updates to the exception handling table and the debug information.
We choose to use this linear scan based algorithm mainly due to its simplicity to implement and debug, 
its linear time complexity, and that it exhaustively covers all instruction sequences of selected lengths.

\input{figs/outline_sequence}

\begin{algorithm}
\caption{General Sequence Outlining}
\label{alg-general-outline}

\SetKwFunction{Visibility}{ComputeVisibility}
\SetKwFunction{Hash}{hash}
\SetKwFunction{Sort}{sort}
\SetKwFunction{VerifySequences}{verifySequences}
\SetKwFunction{CreateOutlineFunc}{createOutlineFunc}
\SetKwFunction{CollectCallSites}{collectCallSites}
\SetKwFunction{CreateBranchingLogic}{createBranchingLogic}
\SetKwFunction{UpdateBranchTarget}{updateBranchTarget}
\SetKwFunction{UpdateMetadata}{updateMetadata}

\SetKw{Continue}{continue}

\newcommand\mycommfont[1]{\small\ttfamily\textcolor{blue}{#1}}
\SetCommentSty{mycommfont}

\KwIn{$ Program $  \tcp*[h]{program to be optimized}  \newline
      $ Length $   \tcp*[h]{longest sequence to consider}  \newline
      $ MinFreq $  \tcp*[h]{frequency threshold}
     }
\KwOut{$ optimizedProgram $}

\BlankLine
\tcp{Step 1: collect potential outline sequences}

\ForEach{$function \in Program$} {

  $visibleSet \longleftarrow \Visibility{function}$

  \For{$len \gets 2$ \KwTo $Length$} {
    \tcp{$sequence$ is of length $len$}
    \ForEach{$sequence \in function$} {
      $skip$ = False

      \ForEach{$inst \in sequence$} {
        \lIf{$inst \in visibleSet$} {
          $skip$ = True
        }
      }
      \lIf{$skip$} {
        \Continue
      }

      $h \gets \Hash{sequence} $

      \tcp{map hash value to frequency}
      $hashToFrequency[h] += 1$
      
      \tcp{map hash value to list of its callsites}
      $hashToCallSites[h].append(sequence)$
    }
  }
}
\tcp{byte-wise identity check on sequences with the same hash value}
$\VerifySequences{hashToFrequency, hashToCallSites}$

\BlankLine
\tcp{Step 2: make outline decisions}

\tcp{sort sequences by length and frequency}
$\Sort{hashToFrequency}$

\ForEach{$hash \in hashToFrequency$} {
  \If{$hashToFrequency[hash]>MinFreq$} {
    $outlineDecisions.append(hashToCallSites[hash])$
  }
}

\BlankLine
\tcp{Step 3: create outlined functions}

\ForEach{$outlineInfo \in outlineDecisions$} {
  \tcp{create new function with outlined sequences}
  $\CreateOutlineFunc{outlineInfo[0]}$
}

\BlankLine
\tcp{Step 4: modify original functions}

\tcp{find out all functions that need updates and the corresponding instructions to be outlined}
$callSites \gets \CollectCallSites{hashToCallSites}$
\ForEach{$CS \in callSites$} {
 
  \ForEach{$outlineSequence \in CS$} {
    \tcp{replace to-be-outlined sequences with branching logic}
    $\CreateBranchingLogic{outlineSequence}$
  }

  \tcp{update branch target indices for control flow instructions}
  $\UpdateBranchTarget{CS}$

  \tcp{update  metadata (i.e., exception handling table) of the function}
  $\UpdateMetadata{CS}$
}

\end{algorithm}

Figure~\ref{fig:outline_sequence} shows three examples of instruction sequences
that are outlined in one of our applications, including a sequence of data 
movements between registers, calling an Objective-C runtime function,
and calling a system function.
Each of the three sequences appeared more than 500 times in the application.

\stitle{Update branch targets.}
Since outlining modifies the semantics and positions of instructions, we need 
to update the relevant control flow instructions for correctness. This includes
both direct and indirect branches.

For direct branch instructions where the branch offset is hardcoded in the 
instruction's encoding, we first identify the target instruction before 
outlining by decoding the offset value. 
During transformation, we record the mapping of the instruction indices before 
and after outlining, so that we can update the branch offset values 
to point to the correct targets when we write out the instruction sequences after outlining.

For indirect branches where the targets are encoded as data-in-code (e.g., jump tables), 
we currently skip the outlining optimization altogether on the particular function.
Modifying the function without updating the content of data-in-code would lead to incorrect logic. 
In \texttt{ld64} terms, we identify such functions by looking for
\texttt{kindDataInCode} type of fixups in an atom. 
We empirically observe that less than 2\% of the overall functions in our applications 
contain data-in-code components, thus skipping such functions has little impact 
on the overall size reduction.
Alternatively, one could parse these data-in-code components and update their 
contents based on changes made by the outliner.

For indirect branches where the targets are expressed as linker relocations, 
since our outlining pass is scheduled before atom ordering and fixup resolution, 
such relocations are symbolically represented during our passes. 
Thus, we do not need to explicitly update such symbolic relocations. 
Instead, we rely on linker's downstream fixup resolution step to correctly 
replace the symbolic fixups with the final addresses of the corresponding functions.

\stitle{Update exception handling table.}
The exception handling table (EHT) describes the program behavior
when an exception happens, including the execution of exception handling code and
any cleanup or stack unwinding actions.
The EHT is encoded in a language-specific data area of the binary.
In each code segment where an exception can potentially happen, the EHT describes both the 
landing pad location (if any) and the required actions 
(e.g., calling an object destructor).
The EHT refers to instructions in the code segments using their relative indices 
within the function they reside in.
As a result, when we modify the code sequences during outlining, we also need
to update the content of the EHT to reflect the modification.
To this end, we provide a \emph{parser} that parses an EHT, and a \emph{rewriter}
that updates the parsed EHT based on the changes made by the outliner.

\input{figs/dwarf_flow}

\stitle{Update debug information.}
Debug information (e.g., in DWARF format~\cite{dwarf}) is crucial 
when running a debugger and analyzing crashes. Debug information allows mapping 
an instruction's address to its original location in the source code.
In a typical compile and link flow, such as the one for iOS applications,
the pre-linking debug information is firstly generated by the compiler and stored
as part of the object file.
Then at link time, the linker generates a debug map that records
the mapping of each function to its final address.
Finally, the debug information linker (e.g., DWARF linker) utilizes the 
debug map to collect the debug information from each object file and link them
to the final DWARF file.
Figure~\ref{fig:dwarf_flow} shows a high-level view of the DWARF linking process.

One important assumption of the above-mentioned DWARF linking flow is that  
functions in the final linked binary are identical to their counterparts in the 
pre-linking object files. In other words, the linker should not modify the content
or the size of the functions.
However, since our linker outliner modifies the functions to reduce their sizes, 
this assumption is no longer valid.
Running the linker outlining pass without updating the DWARF information
would lead to inconsistencies between the executable and the debug information.

We customized the DWARF linker (i.e., the \texttt{dsymutil} tool)
to account for the changes made by the outlining pass.
To achieve this, the \texttt{ld64} linker writes a lightweight auxiliary file containing 
the information on how the outliner modified the content of the functions.
Then the customized DWARF linker reads this file and adjusts the address mapping
based on modifications made by the linker outliner.
We provide a bundled executable linker with the customized DWARF linker for 
building the application.

\subsection{Frame Code Outlining}

Many calling conventions require the callee to explicitly save and 
restore callee-saved registers within prolog and epilog regions of a
function.
Such logic, also known as the frame code, is highly regular and can 
consume non-trivial space.
Linker frame code outlining aims at extracting such code regions into shared functions 
and replacing the original frame code sequences with calls to these shared functions.%
\footnote{We note that a prior work~\cite{fb-pgo} proposed a frame code optimization
technique during LLVM machine IR optimization, while we conduct frame code outlining during linking.}
Figure~\ref{fig:framecode_a} shows an example of a prolog frame code that first
updates the stack pointer, then stores callee-saved registers (\texttt{x19} to \texttt{x30}) on the stack,
and finally updates the frame pointer register \texttt{x29}.

Unlike the general sequence outlining discussed in Section~\ref{sec-general-outline},
frame code sequences are more regular. 
We specialize our optimization to better utilize the regularity.
For example, prolog (epilog) code only appears at the beginning (end) 
of a function.
When optimizing the prolog segments, there is no need to preserve the current 
values of temporary registers since they do not contain live values when
execution enters the current function.
This is especially useful since we can use the temporary registers to store return addresses when
calling the outlined prolog sequences.
Similar reasoning is also applicable to optimize the epilog segments since 
no temporary registers will be live after the epilog sequence.

\input{figs/framecode}

\stitle{Normalize stack offset value.}
A compiler can reduce the number of stack offset adjustment
operations by merging the offset adjustment due to callee-saved registers with
those related to other temporary variables on the stack. 
In Figure~\ref{fig:framecode_a}, the prolog reserves a total of 320 bytes of space 
in the stack for the current function, among which 96 bytes are reserved for the
callee-saved registers, and the remaining 224 bytes for other temporary variables
declared in the current function.
In our frame code outlining pass, we add a normalization step to separate these
two sources of stack offset adjustments.
Figure~\ref{fig:framecode_b} shows the normalized sequence, where we moved the 
stack pointer adjustment to the end of the sequence and updated the constant 
offset values accordingly.
As a result, the normalized frame code sequences from different functions can
share the same outlined framecode function if they store/load the same 
number of callee-saved registers.
This normalization step reduces the number of outlined functions that need to be created, 
thus improving the total size saving.

In this step, we constrain the transformation so that we only write temporary 
variables within the stack's red zone (e.g., 128 bytes for AArch64 architecture~\cite{apple-abi}).
The application binary interface ensures that these locations within the red zone 
will not be modified by other parts of the system.
For platforms without a defined stack red zone, we skip this normalization step to ensure safety.

\subsection{Safe Identical Code Folding}

While outlining explores deduplication at the instruction level, identical code
folding (ICF) reduces code size by discovering functions with identical implementations and replacing them with a shared implementation.
The original \texttt{ld64} linker includes a code deduplication pass where only functions with
the ``autohide'' property are considered.
This limits deduplication's applicable scope because the pass relies on the compiler to explicitly mark qualified functions.
We find that there are a large number of symbols (especially private symbols) 
that are not marked as ``autohide'', leaving significant room for improvement.
We thus customize the deduplication pass in \texttt{ld64} by considering all non-global symbols
as candidates for ICF.

Figure~\ref{fig:icf_sequence} shows an example of three identical functions that
can be merged to reduce the code size.
These are commonly setter and getter methods of variables, which can be either 
manually written or automatically synthesized in languages such as Objective-C.
These three functions are from entirely independent modules, but they all implement
the same logic that stores one byte of data into memory.
In one of our applications, there are more than 2,000 such functions with the exact
two-instruction sequence across different compilation units, making them 
ideal candidates for ICF.

\input{figs/icf_sequence}

\stitle{Check identical functions.}
The ICF pass first computes a hash for every non-global function using the 
function hashing technique described in Section~\ref{sec:analyses}.
It then groups the functions with the same hash value and checks whether folding
can reduce the overall code size.

\stitle{Handle function pointer comparisons.}
A function pointer stores the start address of a function in the memory.
In many languages, such as C, C++, or Objective-C, the programmer or the
runtime is allowed to conduct arithmetic operations over function pointers.
One of the most used arithmetic operations is equality
comparison. It checks whether two function pointers point to the same address
(i.e., the function's implementation).
Figure~\ref{fig:func_ptr_compare_a} shows a toy example of using function pointer 
comparison, where we assume that the implementations of both \texttt{func1} and 
\texttt{func2} are identical.

A straightforward ICF implementation (commonly known as the \texttt{icf=all} option in
modern linkers) directly replaces the duplicate's implementation address with
the address of the other function. 
Such a transformation would lead to an incorrect return value of \texttt{1} for this example.
To ensure safety under function pointer comparison, we improve the \texttt{ld64} linker by implementing
the \texttt{icf\_safe} option.%
\footnote{The \texttt{icf\_safe} option is available in some other linkers~\cite{tallam2010safe} with potentially different implementations. 
Our work introduces an implementation in the \texttt{ld64} linker.}
The \texttt{icf\_safe} option adds redirection logic to preserve correctness under function pointer equality comparison.
Figure~\ref{fig:func_ptr_compare_b} illustrates the assembly code generated by \texttt{icf\_safe}. 
Instead of directly replacing the implementation address of \texttt{func1} with
\texttt{func2}, we add a single-instruction redirection logic to branch to 
\texttt{func2} inside \texttt{func1}, thus preserving the original behavior under function pointer equality comparison.

The safe ICF incurs an one-instruction overhead compared to the \texttt{icf=all} option.
We empirically observe that the additional branching logic introduced by the 
\texttt{icf\_safe} pass does not have any visible performance impact across the applications we tested compared to the baseline version without ICF enabled.

\input{figs/func_ptr_compare}

%% file: figs/instruction_visibility.tex
\begin{figure}[t]
       \centering
       \includegraphics[width=0.55\columnwidth]{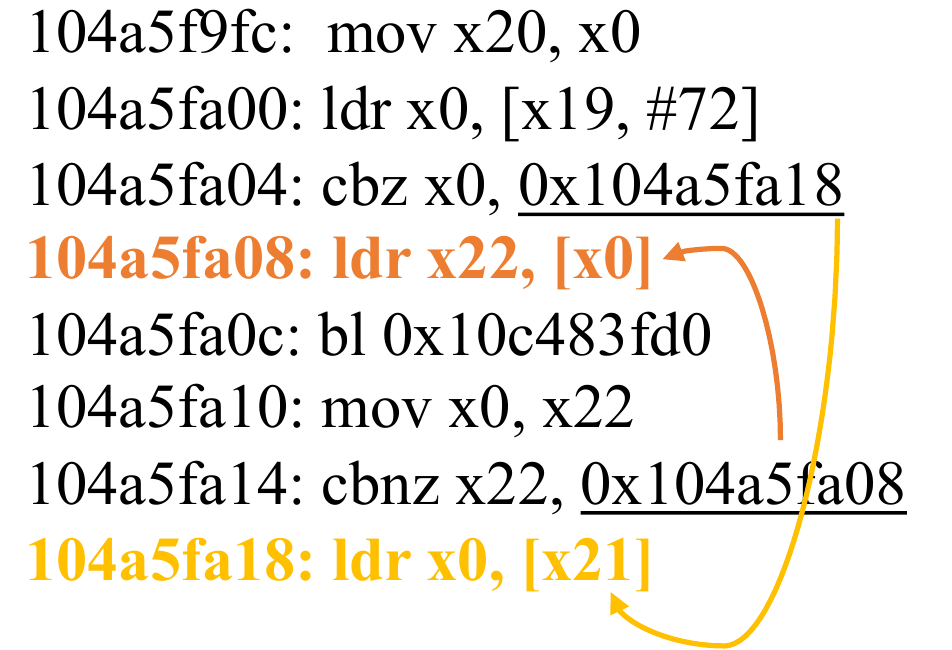}
       \caption{An example of instruction visibility where the two highlighted \texttt{ldr} instructions are the branch targets of the corresponding \texttt{cbz}/\texttt{cbnz} instructions. They are marked as visible.}
       \label{fig:instruction_visibility}
\end{figure}

%% file: figs/outline_sequence.tex
\begin{figure}
     \centering
     \begin{subfigure}[b]{0.22\columnwidth}
         \centering
         \includegraphics[width=0.95\textwidth]{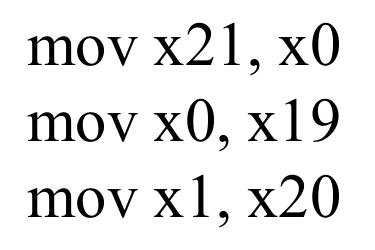}
         \caption{}
     \end{subfigure}
     \hfill
     \begin{subfigure}[b]{0.3\columnwidth}
         \centering
         \includegraphics[width=0.95\textwidth]{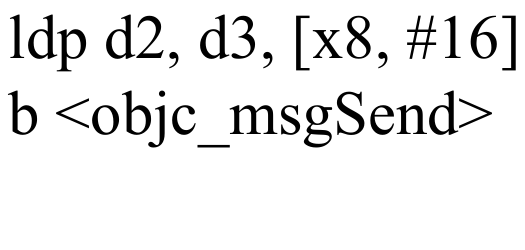}
         \caption{}
     \end{subfigure}
     \begin{subfigure}[b]{0.46\columnwidth}
         \centering
         \includegraphics[width=0.95\textwidth]{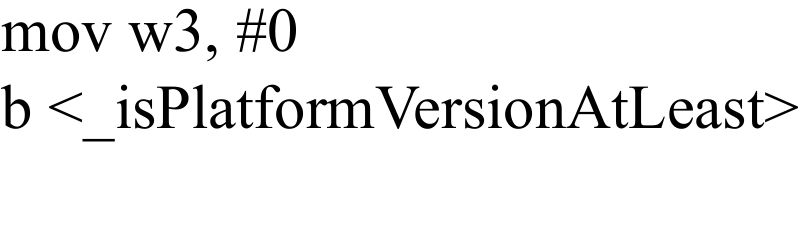}
         \caption{}
     \end{subfigure}
        \caption{Examples of highly repeated sequences as good outlining candidates: (a) data movements between registers; (b) calling Objective-C runtime function; (c) calling system function.}
        \label{fig:outline_sequence}
\end{figure}

%% file: figs/dwarf_flow.tex
\begin{figure}[t]
       \centering
       \includegraphics[width=\columnwidth]{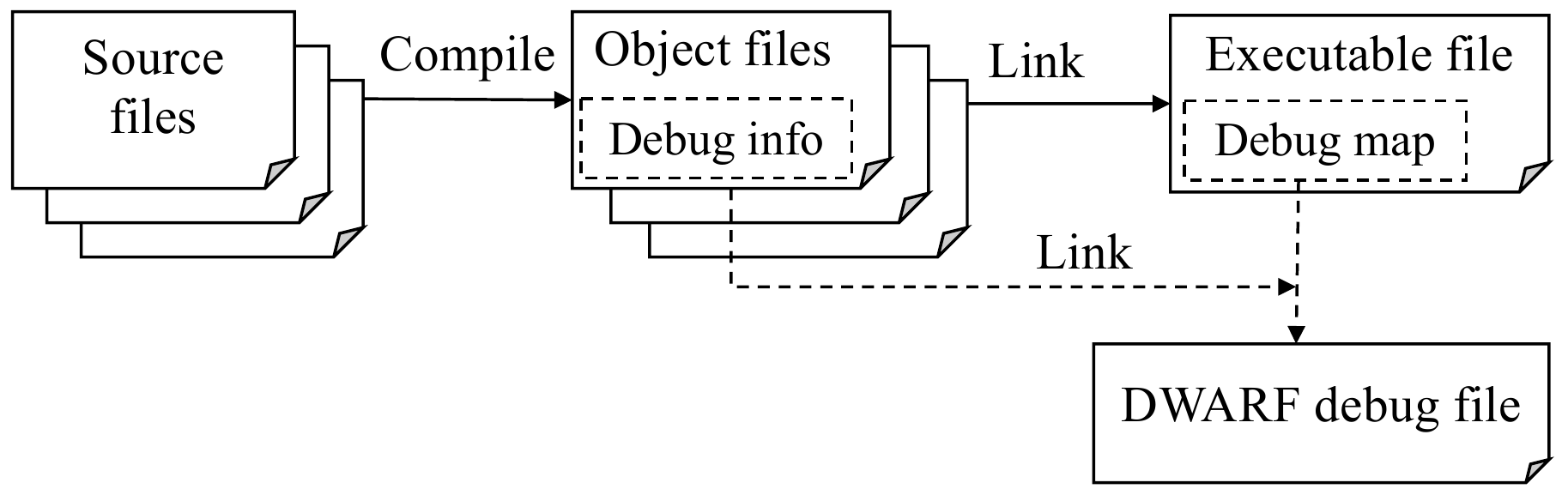}
       \caption{An illustration of the DWARF debug information linking flow.}
       \label{fig:dwarf_flow}
\end{figure}

%% file: figs/framecode.tex
\begin{figure}
     \centering
     \begin{subfigure}[b]{0.48\columnwidth}
         \centering
         \includegraphics[width=0.8\textwidth]{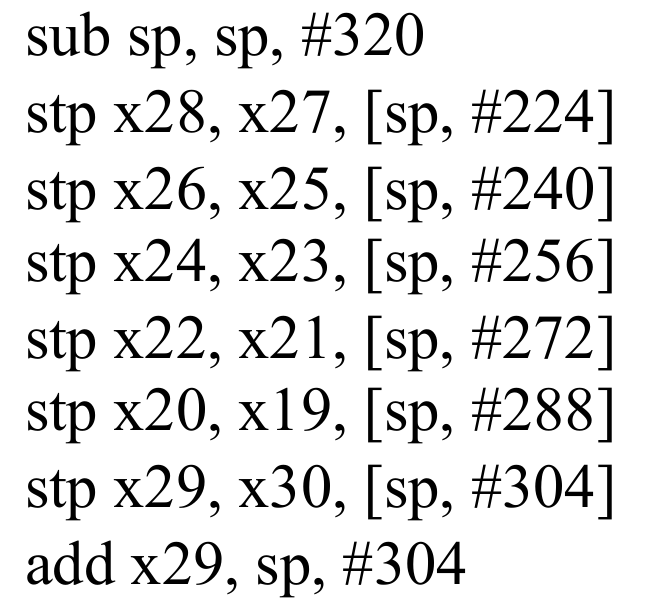}
         \caption{}
         \label{fig:framecode_a}
     \end{subfigure}
     \hfill
     \begin{subfigure}[b]{0.48\columnwidth}
         \centering
         \includegraphics[width=0.8\textwidth]{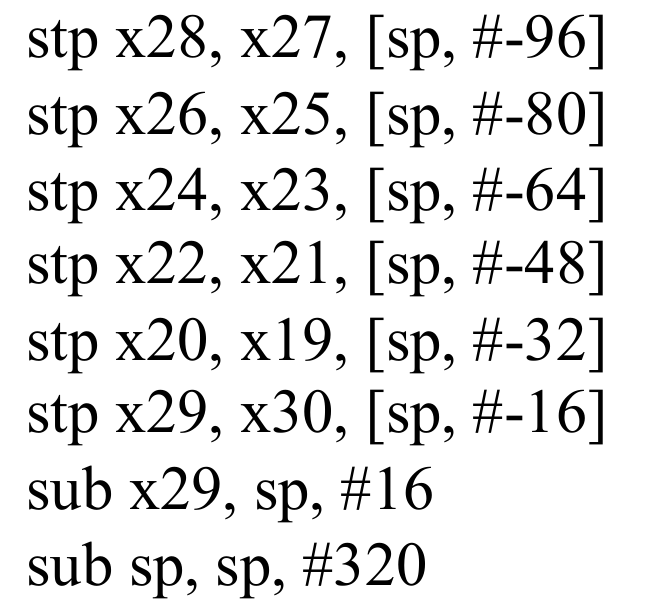}
         \caption{}
         \label{fig:framecode_b}
     \end{subfigure}
        \caption{Frame code sequence and normalization: (a) an example sequence of prolog frame code; (b) the equivalent prolog sequence after normalizing the stack offset.}
        \label{fig:framecode}
\end{figure}

%% file: figs/icf_sequence.tex
\begin{figure}[h]
       \centering
       \includegraphics[width=0.85\columnwidth]{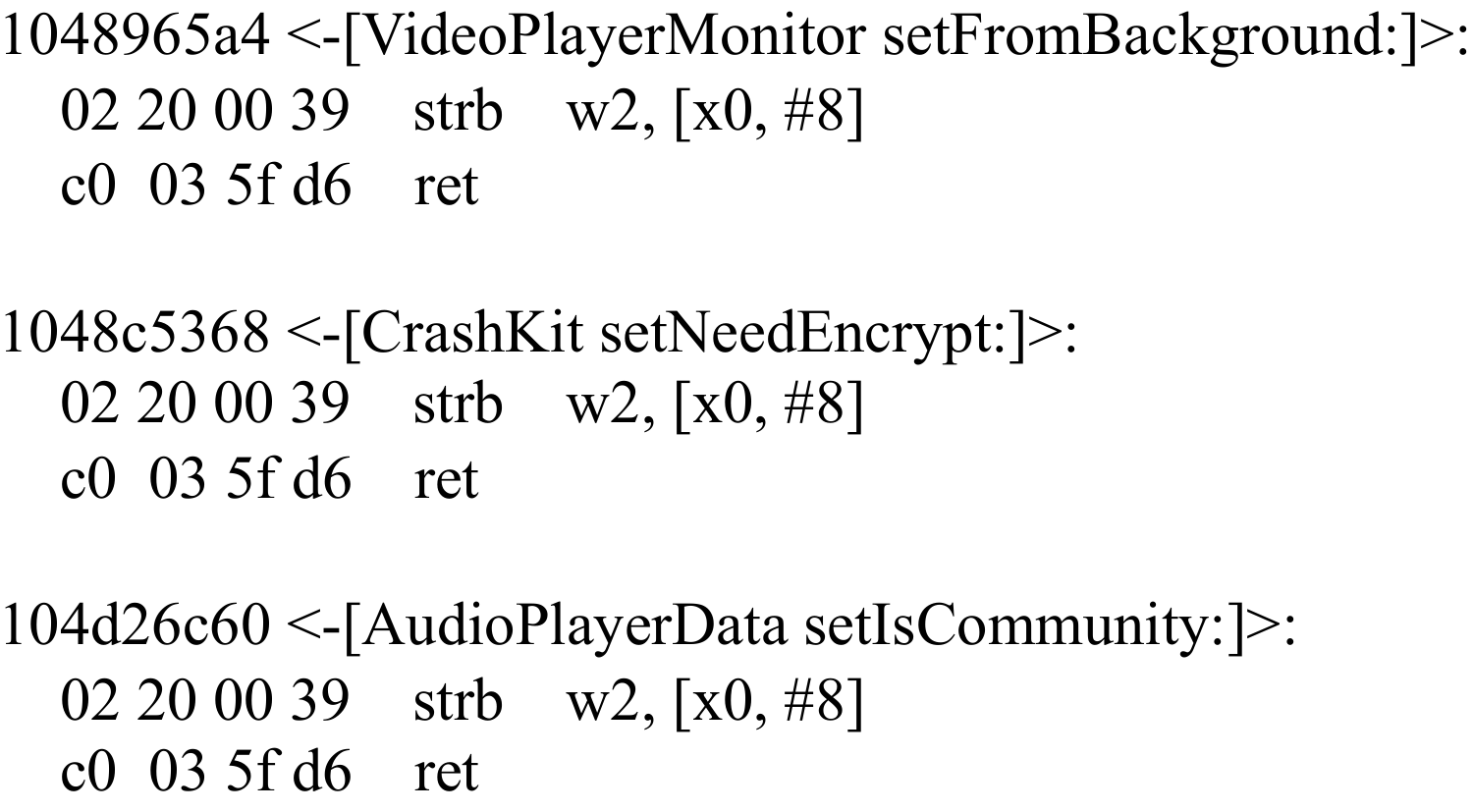}
       \caption{An example of identical functions that can be merged. The function names are modified for confidentiality.}
       \label{fig:icf_sequence}
\end{figure}

%% file: figs/func_ptr_compare.tex
\begin{figure}[t]
     \centering
     \begin{subfigure}[b]{0.49\columnwidth}
         \centering
         \includegraphics[width=0.95\textwidth]{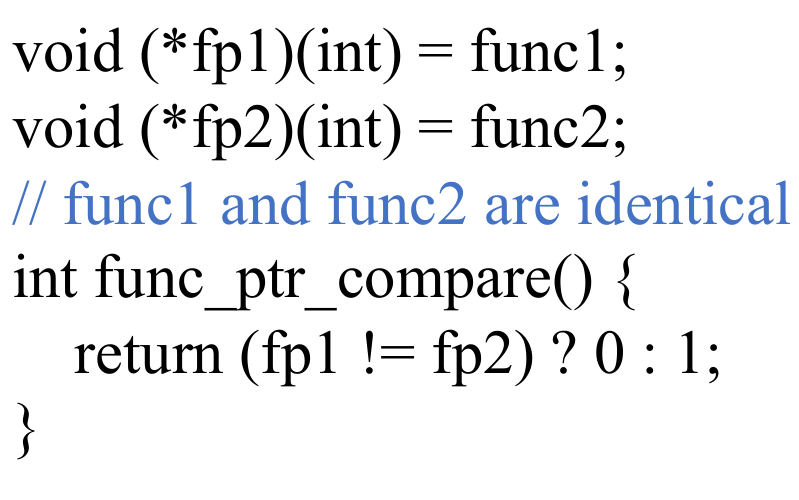}
         \caption{An example of function pointer comparison.}
         \label{fig:func_ptr_compare_a}
     \end{subfigure}
     \hfill
     \begin{subfigure}[b]{0.49\columnwidth}
         \centering
         \includegraphics[width=0.9\textwidth]{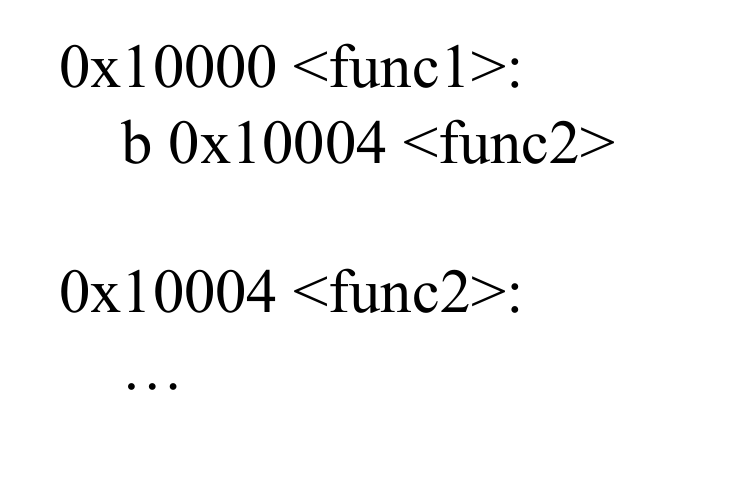}
         \caption{Illustration of assembly code generated by \texttt{icf\_safe}.}
         \label{fig:func_ptr_compare_b}
     \end{subfigure}
        \caption{Handling function pointer comparisons in \texttt{icf\_safe}.}
        \label{fig:func_ptr_compare}
\end{figure}

%% file: experiment.tex
\section{Implementation}
\label{sec:implementation}

\input{figs/size_compare}

We implement previously described ICF and outlining optimizations as additional 
passes in the Apple \texttt{ld64} linker (version 609) ~\cite{ld64}.
Our implementation consists of more than $3,500$ lines of source code, written in the C++ programming language. 
The outline pass includes both the general sequence outlining and the frame code
outlining as discussed in Section~\ref{sec:technique}.
We schedule these new optimization passes in the \texttt{ld64} linker after object file parsing and 
symbol resolution. 
The outlining pass is scheduled before the ordering pass
that determines the total order of all the atoms. This is because the 
outlining pass creates new atoms. Together with those existing ones, the new atoms need to be ordered by the ordering pass.
The safe ICF optimization is implemented as part of the existing code
deduplication pass in \texttt{ld64}, which is scheduled right after the ordering pass. 
The overall \texttt{ld64} linking flow and optimization pipeline are shown in Figure~\ref{fig:overview}.
By default, the optimizations target AArch64 architecture~\cite{arm64-arch} since it is the 
dominating architecture for iOS production devices.
The general optimization principles are equally applicable to other CPU architectures as well.

\section{Experiments}
\label{sec:experiment}
To evaluate the effectiveness and performance of our optimization passes, we conduct extensive experiments on iOS applications. 

\stitle{Applications.}
We apply our optimizations to three widely used commercial iOS applications.
Each of these iOS applications has hundreds of millions of daily active users 
and covers a wide range of mobile application usage scenarios.
\TouTiao{} is a news recommendation application, which provides personalized
text, audio, and video content to end users.
\TikTok{} is a mobile short video hosting and sharing application. It hosts a variety of 
user-created short videos lasting between tens of seconds and a few minutes. 
It also personalizes users' video feed using machine learning based
recommendations.
Lastly, \Lark{} is an iOS client of an enterprise collaboration platform with services 
covering email, instant messaging, video/audio conferencing, audio transcription, online documentation, and so on.
All these applications are written using a mixture of Objective-C, Swift, and C++.
One of the applications contains components written in Rust.
The number of functions in each application ranges from $1\times 10^6$ to $2\times 10^6$ functions.


\stitle{Experimental setup.}
To build the above-mentioned commercial applications, we perform build on an idle build machine with a 2.6 GHz 6-core Intel CPU and 64 GB of RAM running MacOS.
The applications are compiled with Apple's Xcode toolchain (version 13.0) ~\cite{app-xcode} and linked using the \texttt{ld64} linker with our custom passes.

Next, we discuss our experiment details in terms of code size reduction and build time comparison. 
Furthermore, we compare performance impacts with and without our optimizations.  

\subsection{Code Size Reduction}
\label{sec:commercial-app-size}


Figure~\ref{fig:size_compare} shows the size reduction for each application 
when enabling the two optimization passes.
The compiler flags in the baseline have been tuned manually on a per-module 
basis to satisfy the module's specific quality goals independent of our work. 
For example, video editing/playing related
modules are compiled with \texttt{-O3} for maximum performance, while other less
computationally intensive modules are compiled with \texttt{-Oz} or \texttt{-Os} to 
minimize code size. 
In this experiment, we do not change any existing compiler flag that the build system is already using. 
Instead, we only add the \texttt{icf\_safe} and \texttt{outline} linker flags. 
This helps to ensure a fair comparison.

The size reduction is measured in two ways. Binary size measures the size of a
binary file produced by a linker. It can be either a library or an executable. A binary file is commonly composed
of multiple sections including both text section(s) and data section(s). Our optimization
passes only operate on the text section(s) inside a binary file.
The size of an iOS app store package (i.e., IPA) refers to the size of an application package 
delivered through the Apple App Store~\cite{app-store}.
It contains both binary files and resource files, including images, language
support files, resource bundles, and so on.
The IPA file is usually downloaded in a compressed format and then decompressed during installation.

\begin{table*}
\centering
\small
\caption{Build time profile on \TouTiao{}, \TikTok{} and \Lark{} and the percentage overhead of our passes. m: minutes, s: seconds.}
\begin{tabular}{lrrrr}
\toprule
                                 & \TouTiao{} & \TikTok{} & \Lark{} & Geomean \\
\midrule
Setup and compile                & 39m8s   &  26m47s &  18m57s&  -- \\
Link                             & 13m23s  &  22m42s &  9m12s &  -- \\
\quad  - \texttt{icf\_safe} pass & 45s     &  1m35s  &  22s   &  -- \\
\quad  - \texttt{outline} pass   & 6m15s   &  8m21s  &  4m24s &  -- \\ 
Total build time                 & 52m31s  &  49m29s &  28m9s &  -- \\
\midrule
\texttt{icf\_safe} build time overhead         & 1.4\%   & 3.2\%  & 1.3\%  & 2.0\% \\
\texttt{outline} build time overhead           & 11.9\%  & 16.9\% & 15.6\% & 14.8\% \\
\texttt{icf\_safe} + \texttt{outline} build time overhead & 13.3\%  & 20.1\% & 16.9\% & 16.7\% \\
\bottomrule
\end{tabular}
\label{tbl:link_time}
\end{table*}

Across all three applications, the two optimizations combined achieved 18.4\% size 
reduction in uncompressed binary, and 4.3\% size reduction in the IPA file on average.
Specifically, the ICF pass reduces the size of the uncompressed applications 
by 4.4\% and the linker outlining pass further reduces the uncompressed 
size by 14.0\% on average. We observe that our approach achieves more significant size
savings in both the binary size and the IPA size than start-of-the-art
LTO based approaches~\cite{uber-lto, fb-pgo}. 
In addition, our results agree with~\cite{fb-pgo} in that the reduction in compressed 
IPA size is much smaller than that in binary size. 
This is likely because the IPA compressor is more effective on binaries than other resources.

\subsection{Build Time Comparisons}

In our build configuration, the two size optimization passes are enabled only in 
release build, and debug build is entirely intact.
To measure the link time impact, we profile the link time increase due to the two passes.

Table~\ref{tbl:link_time} shows the build time breakdown for the three applications and 
link time consumed by the two passes individually.
The total build time is the time taken from setting up the build environment to 
the linker finished producing the final IPA file. The build process includes cloning source repositories, 
compiling source files individually to produce object files, and linking them together to produce the final images.
We observe that the \texttt{icf\_safe} pass increases the link time by 2.0\% on average,
while the \texttt{outline} pass accounts for 14.8\% of the link time across the three applications.
We observe that the build time is vastly dominated by time spent compiling source files.
Overall, the two passes collectively incur a 16.7\% increase in the overall build time.
These two passes are enabled only in the release stage and do not impact application
developers' feature development workflow.
The increase in build time is mild and considered acceptable by the applications' build teams.

\begin{table}
\centering
\small
\caption{Application startup time comparison with \TouTiao{}. Baseline: results with default flags; Size Optimized: results with \texttt{icf\_safe} and \texttt{outline} on. The numbers are averaged over five runs.}
\begin{tabular}{lcc}
\toprule
                        & Baseline (ms) & Size Optimized (ms) \\
\midrule
Library loading         & 269.0    & 254.5  \\
Object loading          & 205.6    & 213.7  \\
Application Launching   & 240.6    & 241.8  \\
Initial Frame Rendering  & 107.0    & 112.9  \\
\midrule
Total                   & 822.2    & 822.9  \\
\bottomrule
\end{tabular}
\label{tbl:perf_compare}
\end{table}

\subsection{Performance Impact}

Since both \texttt{icf\_safe} and \texttt{outline} passes introduce additional control flow
instructions into the optimized code, it is important that they do not cause
performance degradation.
Here we compare several key performance metrics in mobile applications.
The experiments are conducted on an iPhone SE2 with six CPU cores, 
3 GB of RAM, and a quad-core GPU.
We focus on \TouTiao{} since it contains a diverse set of use
scenarios commonly found in mobile applications.
The performance evaluation is conducted by appropriate profiling schemes provided by Apple's Xcode toolchain~\cite{app-xcode}.

\stitle{Application startup time.}
Application startup time is one of the key performance metrics for mobile
applications. It measures the time between a user clicking on an 
application icon and the application finishing displaying its first frame after rendering.
The startup delay directly translates into the wait time before a user can interact with the application.
Reducing the startup delay can have a direct impact on user experience and significantly improve user engagement~\cite{yan2012fast}.

Table~\ref{tbl:perf_compare} presents startup delay impacts from our size optimizations on \TouTiao{}.
The startup time is divided into the following four sequential phases:
(i) library loading measures the time for an application to load the system's dynamic libraries;
(ii) object loading measures the time for the application to load its initial objects;
(iii) application launching includes logic, I/O, and memory access to set up various components and contents in an application, and
(iv) initial frame rendering draws the first frame on the screen which also marks the completion of the startup process.

For both the baseline and optimized versions, we collect the average results after five identical runs.
It is clear from the table that the size optimizations have a negligible impact on the startup time.
This is mainly because the mobile applications we presented are mostly I/O bound, and
the potential overhead caused by the new passes does not cause user-visible impact based on our profiling results.

\begin{table}
\centering
\small
\caption{Video playing frames-per-second (FPS) comparison in \TouTiao{}. Baseline: results with default flags; Size Optimized: results with \texttt{icf\_safe} and \texttt{outline} on.}
\begin{tabular}{lcc}
\toprule
       & Baseline (FPS) & Size Optimized (FPS) \\
\midrule
Run 1   & 36.58   & 37.15     \\
Run 2   & 36.00   & 35.41     \\
Run 3   & 37.20   & 37.16     \\
\midrule
Average & 36.59   & 36.57     \\
\bottomrule
\end{tabular}
\label{tbl:video_fps}
\end{table}

\stitle{Video playing performance.}
Video playing is another important use case in mobile applications.
To measure the impact of our size optimizations on video playing performance,
we collect the frames-per-second (FPS) metric using the video feed page in \TouTiao{}. 
For both the baseline and the optimized versions, we measure the average FPS
over three one-minute long video play sessions using an automated script, where the script 
switches to the next available video every three seconds.
The video feed algorithm randomly selects from a pool of available videos with similar characteristics.
Table~\ref{tbl:video_fps} shows the results of three runs and their average.
We observe that both the baseline and our optimized versions have indistinguishable FPS numbers, indicating
that the size optimization's impact on \TouTiao{}'s video playing is negligible.

%% file: figs/size_compare.tex

\begin{figure*}
     \centering
     \begin{subfigure}[b]{\columnwidth}
         \centering
         \includegraphics[width=\textwidth]{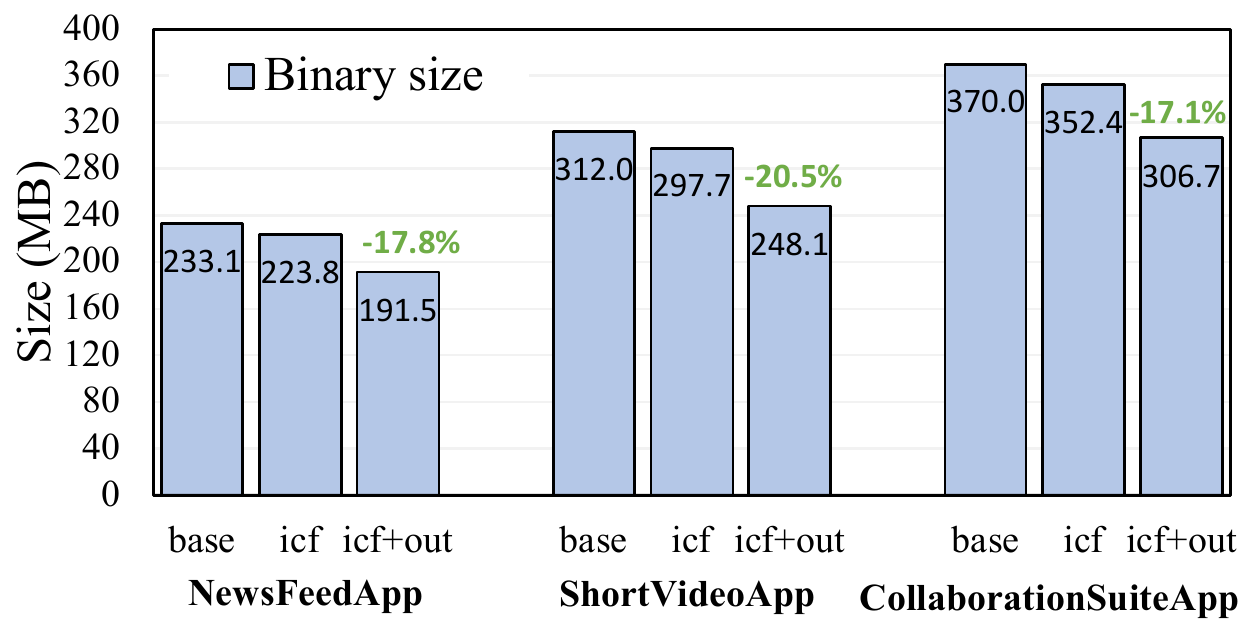}
         \caption{Binary size: size of the uncompressed binary executable.}
         \label{fig:size_compare_1}
     \end{subfigure}
     \begin{subfigure}[b]{\columnwidth}
         \centering
         \includegraphics[width=\textwidth]{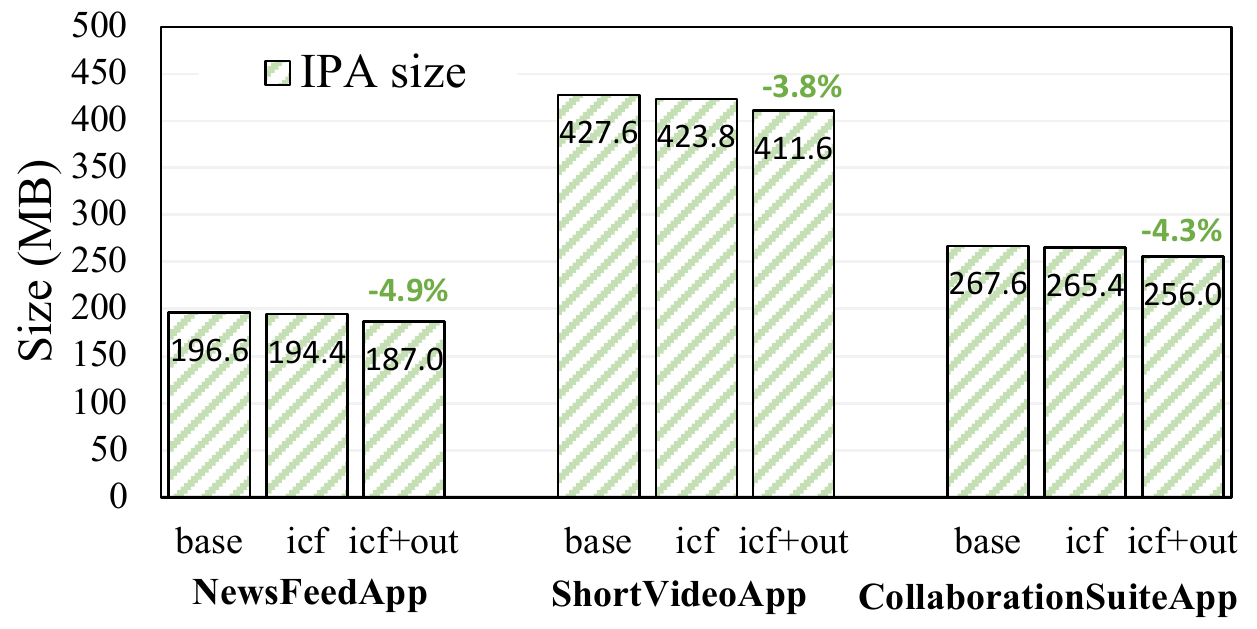}
         \caption{IPA size: download size of the compressed application package.}
         \label{fig:size_compare_2}
     \end{subfigure}
        \caption{Size optimization for \TouTiao{}, \TikTok{} and \Lark{}. \texttt{base}: baseline approach with no linker size optimizations. \texttt{icf}: enable the safe ICF optimization. \texttt{icf+out}: enable both safe ICF and linker outlining.}
        \label{fig:size_compare}
\end{figure*}

%% file: related.tex
\section{Related Work}
\label{sec:related}

Our work is closely related to work in compiler optimizations for code size reduction, link time and 
post-link-time optimizations, linker improvements, and optimizations for mobile applications.

\stitle{Code size optimization.} 
Rodrigo et al.~\cite{functionmerge1,functionmerge2} explored various algorithms for finding similar functions and merging them to reduce application size. They use hash code to check function similarity which we also use to find identical functions. Lee et al. ~\cite{fb-pgo} and Chabbi et al. ~\cite{uber-lto} perform code size optimizations for commercial iOS applications at the machine IR level during compilation, and they find that machine IR is a better target level than LLVM IR for their work. However, we perform optimizations on the machine instructions during linking. As a result, our technique reduces the build overhead. 

More recently, machine learning-based approaches, such as MLGO~\cite{trofin2021mlgo} and CompilerGym~\cite{compilergym-cgo22} proposed to utilize machine learning models instead of heuristics. A combination of these approaches with our work might yield additional benefits in size reduction. Superpack~\cite{superpack-fb} uses a custom compression algorithm to reduce Android bytecode~\cite{dex-bytecode} size. 
In contrast, this work focuses on native code using code transformations instead of compression. 
Moreover, this work complements several compiler optimizations~\cite{thomas1971catalogue, torczon2007engineering, gcse1970, kennedy1999, wegman1991, rosen1988global, lau2003, ernst1997,edler2014,chen2003code, massalin1987superoptimizer} to further reduce the code size.

\stitle{Link time and post link time optimization.}   
Glek et al. ~\cite{lto1} developed LTO for GCC~\cite{gcc} for performance and package size reduction. 
However, their approach is memory intensive and has scalability issues.
Johnson et. al.~\cite{johnson-lto} introduces ThinLTO, which is a lightweight LTO scheme that mostly runs in parallel and reduces both build time and memory usage. 
ThinLTO shows performance benefits similar to a full LTO approach, while its build time and memory consumption are smaller than that of the full LTO scheme.
Several post-link-time optimizations~\cite{panchenko2019bolt,llvm-propeller} leverage profile data for 
post-link optimizations for data center applications, while the focus of this work is mobile applications.   

\stitle{Linker improvement.}
Traditional linkers~\cite{gnu-linker1,bsd-linker1,gold-linker1} focused on correctness, robustness, stability, and backward compatibility. 
Since link time is the dominating factor within modern rapid develop-build-debug cycles, newer linkers~\cite{gold-linker1,lld-linker1,mold-linker1} instead focus on reducing the link time by utilizing parallel data structures and algorithms~\cite{mold-linker1}. 
On the contrary, this work focuses on code size reduction through novel program analysis and optimizations conducted inside a linker.

\stitle{Optimization for mobile applications.}
More generally, improvements for mobile applications cover language design and compiler optimizations.
This includes size reduction~\cite{asm,redex}, stalled feature flags removal~\cite{ramanathan20piranha}, 
improving Swift protocols~\cite{barik19swift}, and so on. 
In addition, a large body of literature exists regarding performance optimizations 
for mobile applications, by improving responsiveness~\cite{lin-cocurrency}, memory management~\cite{lebeck2020end}, state management~\cite{farooq2020livedroid}, as well as startup time~\cite{parate2013practical,yan2012fast}. 
Our work explores code size optimization which complements existing work.

%% file: conclusion.tex
\section{Conclusion}
\label{sec:conclusion}

In this work, we propose a novel framework for performing linker code size 
optimization for native mobile applications. It focuses on enhancing the linker with transformation
passes that outline common code sequences and deduplicate identical functions.
We reduce the binary size of three widely-used commercial iOS mobile applications by 
18.4\% on average, without any user noticeable performance degradations. 
Compared to existing LTO-based size optimizations, our work also 
significantly reduces build time overhead by confining the transformations
within the linker, thus avoiding the need to piggyback on the compiler's 
optimization passes which tend to be prohibitively expensive for size optimization.

Future directions include improving the effectiveness of the outlining pass by 
optimizing for language-specific features, adding support for data section
deduplication, using profiles to better guide transformation targets, and porting the 
optimizations to other linkers.

\begin{acks}
We would like to thank Yang Yu, Zhangjing Yuan, Kehong Huang and Chi Zhang 
for their help in testing and integrating this work in the applications.
We thank Yuanshuo Zhu and Luchuan Guo for their support in this project.
We appreciate Justin Wei and the anonymous reviewers' valuable comments 
and suggestions.
\end{acks}